\documentclass[%
 reprint,
 amsmath,amssymb,
 aps,
]{revtex4-2}
\usepackage{graphicx}
\usepackage{dcolumn}
\usepackage{bm}
\usepackage[export]{adjustbox}
\usepackage{subcaption}

\begin{document}

\preprint{APS/123-QED}

\title{
Edge-State-Mediated RKKY Coupling in Graphene Nanoflakes
}

\author{Ahmet Utku Canbolat}
\email{utku.canbolat@fau.de}
 \altaffiliation[Current address:  Institute for Multiscale Simulation, Friedrich-
Alexander-Universität, Cauerstraße 3, 91058 Erlangen, Germany\\
Competence Unit for Scientific Computing, CSC, Friedrich-
Alexander-Universität, Martensstraße 5a, 91058 Erlangen, Germany.]{}
\author{\"Ozg\"ur \c{C}ak{\i}r}%
 \email{ozgurcakir@iyte.edu.tr}
\affiliation{%
 Department of Physics, Izmir Institute of Technology, 35430 Urla, Izmir, Turkey
}%

\date{\today}

\begin{abstract}
We investigate the long-range behavior and size dependence of the Ruderman-Kittel-Kasuya-Yosida (RKKY) interaction in hexagonal and triangular graphene nanoflakes with zigzag and armchair edges. We employ the tight-binding model with exact diagonalization to calculate the RKKY interaction as a function of the distance between magnetic impurities, nanoflake size, and edge geometry. Our findings demonstrate a strong dependency of the RKKY interaction on edge geometry and flake size, with notable changes in the RKKY interaction strength. We further analyze the influence of structural defects on the interaction strength of exchange interactions. 

\end{abstract}

\maketitle


\section{\label{sec:level2}Introduction:\protect}

Graphene nanoflakes, finite-sized graphene structures with dimensions in the nanometer range, exhibit intriguing electronic and magnetic properties that are highly dependent on their shape, size, and edge configuration. Four primary types of graphene nanoflakes have been widely studied in the literature: triangular and hexagonal flakes with armchair and zigzag edges \cite{Kaxiras2008, Kuc2010, Yazyev2012, Deng2019, Singh2014}. The unique electronic and magnetic properties of these flakes are closely linked to their edge termination type, sublattice asymmetry which in turn determine their magnetic behavior\cite{Lieb1989, Dresselhaus1996, Fujita1996}. Indirect exchange interaction between site localized magnetic moments in graphene was shown to be of ferromagnetic(FM)/anti-ferromagnetic(AFM) character depending on the sub-lattice type of the impurities and exhibit a peculiar  
$\sim 1/R^3$ type behavior in neutral graphene due to vanishing DOS at the Fermi energy\cite{Sherafati2011,  Kogan2011}. In doped graphene, exchange interaction decays as $1/R^2$, with an amplitude proportional to the density of states at the Fermi energy, oscillating between FM/AFM character at the Fermi wavelength as a function of distance between the magnetic moments\cite{Sherafati2011a, Kogan2013,Power2013}.   

Indirect exchange interactions and the role of edge states in graphene nanoribbons was investigated in a number of works\cite{Black-Schaffer2010, Sun2013, Zhukov2013}. 
The role edge states in phospherene ribbons was studied by by Islam et al. in \cite{Islam2018}. The role of topological edge states on RKKY interactions was recently by Laubscher et al. in \cite{Laubscher2023}. Electrical control of  RKKY interactions at long distances via  whispering gallery modes was investigated in \cite{Canbolat2019}. The role of DOS and dispersion on RKKY interactions in bilayer graphene was studied by Klier et al.\cite{Klier2014, Klier2016}, and in materials featuring quartic-dispersion was studied in \cite{Canbolat2022}.

The RKKY interaction in graphene nanoflakes was investigated by K. Sza\l{}owski in 2011\cite{Szalowski2011, handbook2016} where they primarily focused on small and fixed-size graphene nanoflakes, explored the influence of modified electronic structure and edge effects on RKKY interactions. \'{A}valos-Ovando et al. studied the nature of exchange interactions in triangular zigzag-terminated MoS$_2$ nanoflakes, revealed that their spatial decay behavior deviates from the 2D behavior and acquires non-isotropic as well as non-collinear character. The influence of external electric fields on RKKY interaction on edges of MoS$2$ nanostructures was studied by Mousavi et al.\cite{Mousavi2021}. 


In this work we address the characteristics of the indirect exchange interactions in triangular and hexagonal graphene flakes with armchair and zigzag edges. We reveal the role of edge states and flake size on exchange interactions.  We employ tight binding method to determine the exchange energy where itinerant electrons are coupled to site-localized moments with a Heisenberg type interaction\cite{Black-Schaffer2010}.  We further investigate the influence of edge defects on the RKKY interaction in graphene nanoflakes.


\section{\label{sec:level1}The Model and The Results\protect}

\begin{figure}
    \centering
    \includegraphics[width=0.9\linewidth]{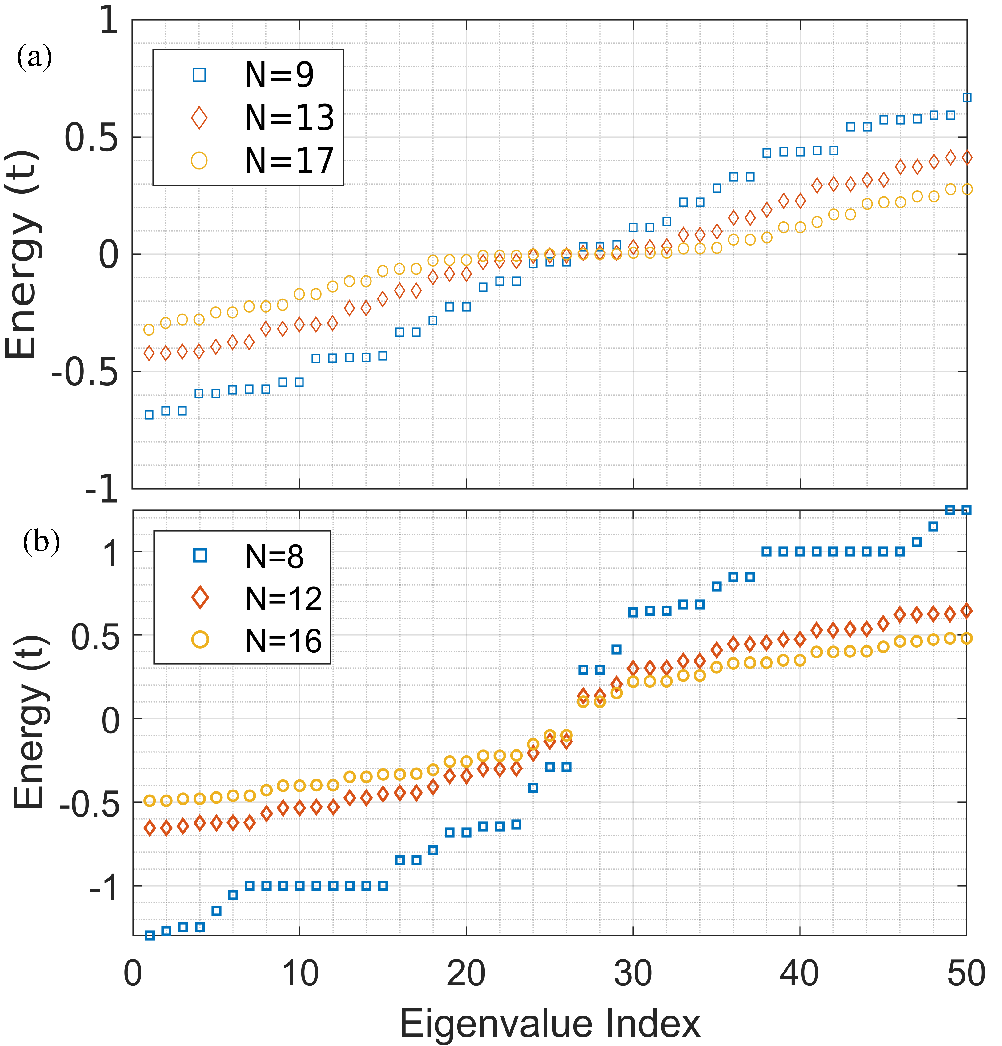}
     \caption{The graph shows the 50 energy eigenvalues around the zero energy for (a) a zigzag edged  hexagonal flake, (b) armchair edged hexagonal graphene flake. Here in the legends $N$ denotes the number of edge atoms.  }
     \label{fig:hexagonal_eigs}
\end{figure}

We considered four different systems for graphene nanoflakes. These are hexagonal and triangular flakes with zigzag and armchair edges. The electronic properties of these systems are well-known \cite{Zarenia2011}. Here, we calculated the electronic properties to show certain features. To calculate the energy, we used the nearest neighbor tight-binding Hamiltonian described as follows:

\begin{align} \label{eq:hamiltonian_tb}
H = -t \sum_{<i,j>, \sigma} c_{i, \sigma} ^\dagger c_{j, \sigma} + \text{h.c.}
\end{align}
where $c_{i, \sigma}$ and $c_{i, \sigma} ^\dagger$ are respectively the annihilation and creation operators at the $i$th site with spin $\sigma$. $t$ is the hopping constant, and its value is approximately 2.8 eV. Fig. \ref{fig:hexagonal_eigs} shows the energy eigenvalues as a function of the eigenvalue index for hexagonal graphene flakes with different sizes. $N$ is the number of atoms on one edge of a flake. The eigenvalue index enumerates the eigenvalues in an ascending order. Since large systems have many eigenvalues, we only plotted around fifty eigenvalues centered at zero-energy for better visualization. 
Figure \ref{fig:hexagonal_eigs}(a) and \ref{fig:hexagonal_eigs}(b) correspond to zigzag and armchair-edged hexagonal flakes, respectively. 
As seen in the figure, the zigzag-edged flakes have zero energy states, and the number of these states increases  with the size of the flake. On the other hand, armchair-edged flakes have a band-gap at zero energy.

\begin{figure}
    \centering
    \includegraphics[width=0.9\linewidth]{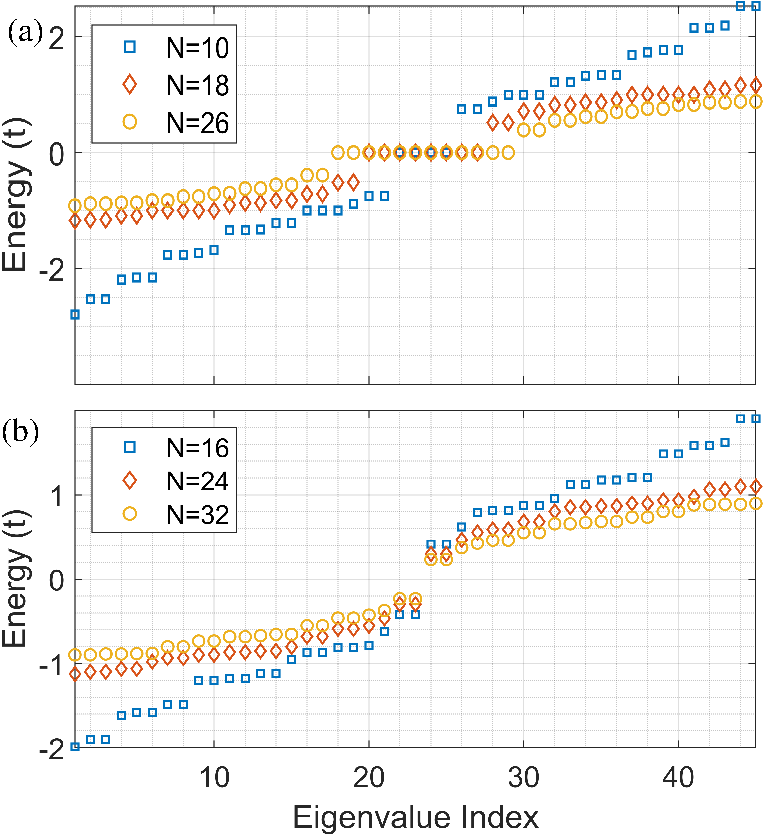}
     \caption{The graph shows the 46 energy eigenvalues around the zero energy for (a) a zigzag edged triangular flake, (b) armchair edged triangular graphene flake. Here in the legends $N$ denotes the number of edge atoms.}
     \label{fig:triangular_eigs}
     \end{figure}

Triangular flakes exhibit a behavior similar to the hexagonal flakes. Figure (\ref{fig:triangular_eigs}) shows that zigzag-edged triangular flakes have zero-energy states whereas armchair edged triangular flakes have band gap at the zero energy. 

\subsection{Magnetic exchange interaction in nanoflakes}

The Hamiltonian of the system can be modeled as the sum of the tight-binding Hamiltonian with the Heisenberg-type spin-spin interaction term as follows:

\begin{align} \label{eq:hamiltonian}
H = -t \sum_{<i,j>} (c_i^\dagger c_j + \text{h.c.}) + J_k \sum_{i=1, 2} \textbf{I}_i \cdot \textbf{s}_i\ 
\end{align}
Here, $ c_i$ and $c_i^\dagger$ respectively indicate the annihilation and creation operators for $i$th site. $t$ is the nearest-neighbor hopping constant. $\textbf{I}_i$ is the localized magnetic moment, and $\textbf{s}$ is the spin of itinerant electrons at site $i$. $J_k$ is the coupling between $\textbf{S}_i$ and $\textbf{s}$. 

RKKY theory states that using the second-order perturbation, one can write the equation (\ref{eq:hamiltonian}) as the interaction between two magnetic moments at sites $i$ and $j$ with an effective coupling constant $J_{ij}$ as follows:
\begin{align}\label{rkky_model}
    H_{\text{RKKY}} = J_{ij}\ \textbf{I}_i \cdot \textbf{I}_j
\end{align}
The RKKY interaction model Eq. (\ref{rkky_model}) describing the magnetic moments by an effective exchange interaction is valid in the weak coupling limit\cite{Ruderman1954, Yosida1957, Kasuya1956}. 

\begin{figure}
    \centering
    \begin{minipage}{0.5\textwidth}
        \centering
        \includegraphics[width=\linewidth]{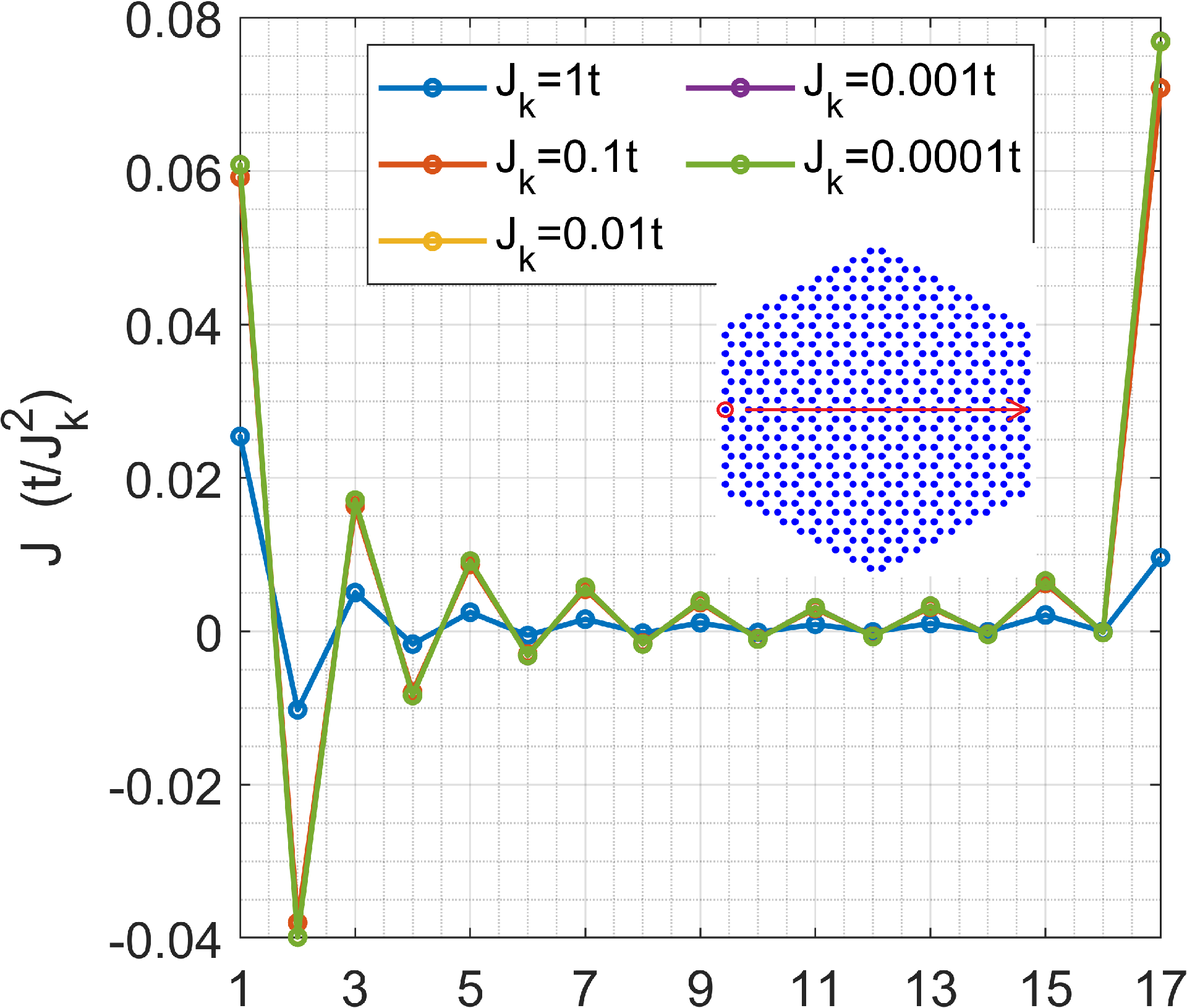}
    \end{minipage}
     \caption{ The graph shows the RKKY interaction strength, on a zigzag-edged hexagonal flake, with respect to the site indicated by the red circle in the inset, where the second site lies in armchair direction in the direction of the red arrow shown in the inset, all the way to the mid-point of the opposite edge. }
     \label{fig:convergence}
\end{figure}

We calculate the effective coupling $J_{ij}$ using exact diagonalization.  The energy difference between ferromagnetic (FM) and anti-ferromagnetic (AFM) configurations is proportional to $J_{ij}$\cite{Black-Schaffer2010, Power2013}. We consider 
magnetic impurities aligned in the same(opposite) direction corresponding to 
FM(AFM) configurations. Then, $J_{ij}$ can be calculated by taking the energy difference between FM and AFM configurations as $J_{ij} = [E(\text{FM}) - E(\text{AFM}]/2$. In Fig. \ref{fig:convergence} (a), the exchange energy is 
calculated for a zigzag-edged hexagonal flake, where one of the magnetic moments is at the center of an edge where the other one is located in a direction towards the center of the opposite edge. Figure \ref{fig:convergence} shows the interaction $J_{ij}$ as a function of distance for different local exchange coupling values $J_k = 10^{-n}t$ with $n=0, 1, 2, 3, 4$. Here it is seen that the behavior of magnetic exchange coupling changes with the value of $J_k/t$. The exchange coupling converges to a characteristic form as the local exchange interaction $J_k$ decreases. For large local coupling $J_k$, exchange coupling becomes short-ranged, albeit edge effects survive. We are interested in the the weak coupling limit and we will assume $J_k = 10^{-5}t$ for the rest of this work. 

In Fig. \ref{fig:convergence} exchange energy is shown for a zigzag-edged neutral hexagonal graphene flake for various local coupling strengths, where one of the magnetic moments is located at the center of an edge and the other one is located in the armchair direction shown by the red line in Fig. \ref{fig:convergence} inset. Here RKKY coupling constant $J_{ij}$ is calculated as a function of distance for different local exchange coupling values $J_k = 10^{-n}t$ with $n=0, 1, 2, 3, 4$. The exchange coupling in units of $J_k^2/t$,  converges to a characteristic form as the local exchange interaction $J_k$ decreases. We are interested in the the weak coupling limit and we will assume $J_k = 10^{-5}t$ for the rest of this work. Exchange energy initially decays with distance and it is of FM(AFM) character, i.e. $J_{ij}<0(J_{ij}>0) 0$ when the magnetic impurities are on same(different) sublattice. However as the second impurity approaches the opposite edge the exchange energy is enhanced for sites of opposite sublattice type.

\begin{figure}[!h]
    \begin{subfigure}{0.2\textwidth}
    \includegraphics[width=\linewidth]{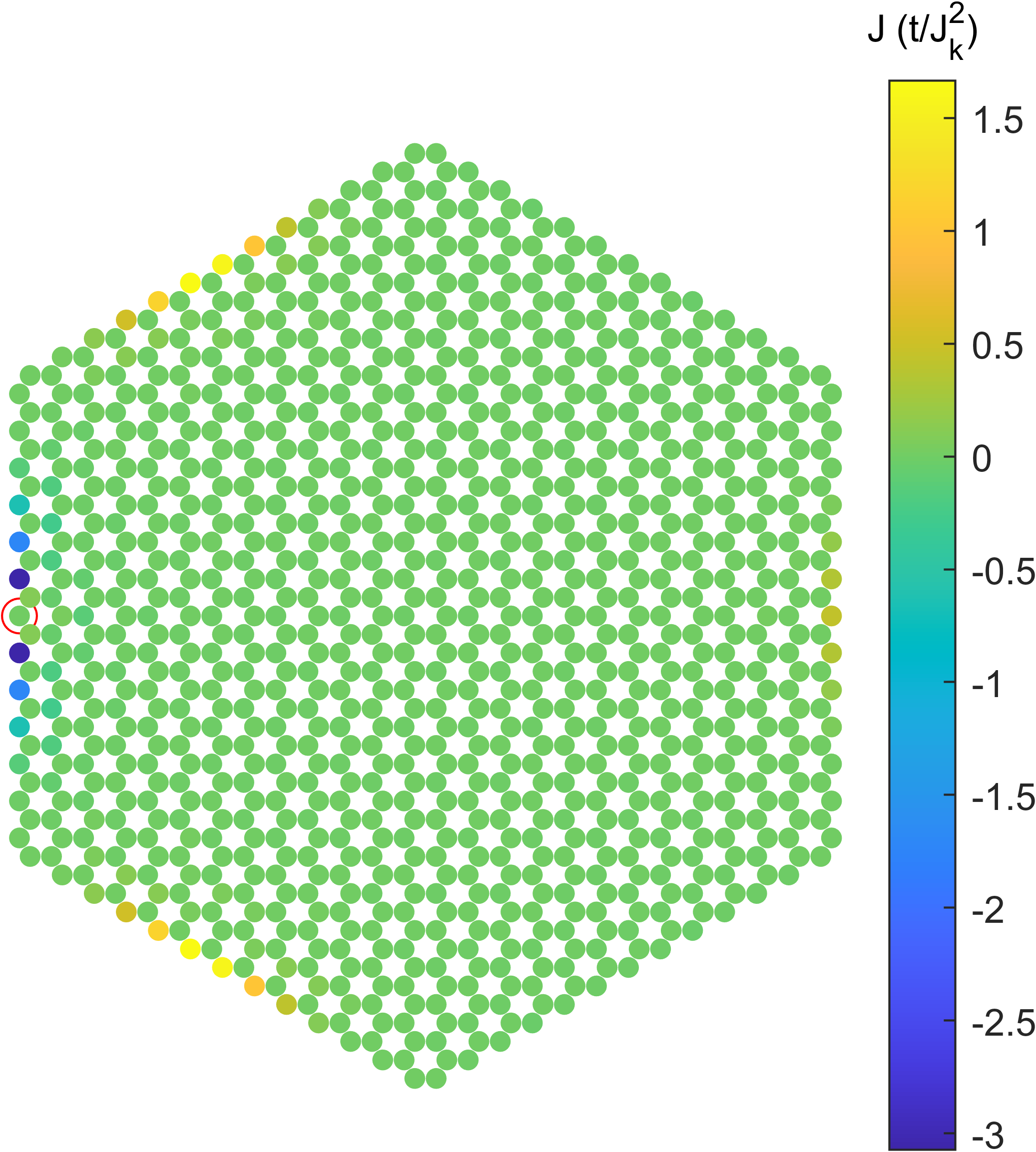}        
    \vspace{-0.8cm}\caption{}
    \end{subfigure}
    \begin{subfigure}{0.2\textwidth}
    \includegraphics[width=\linewidth]{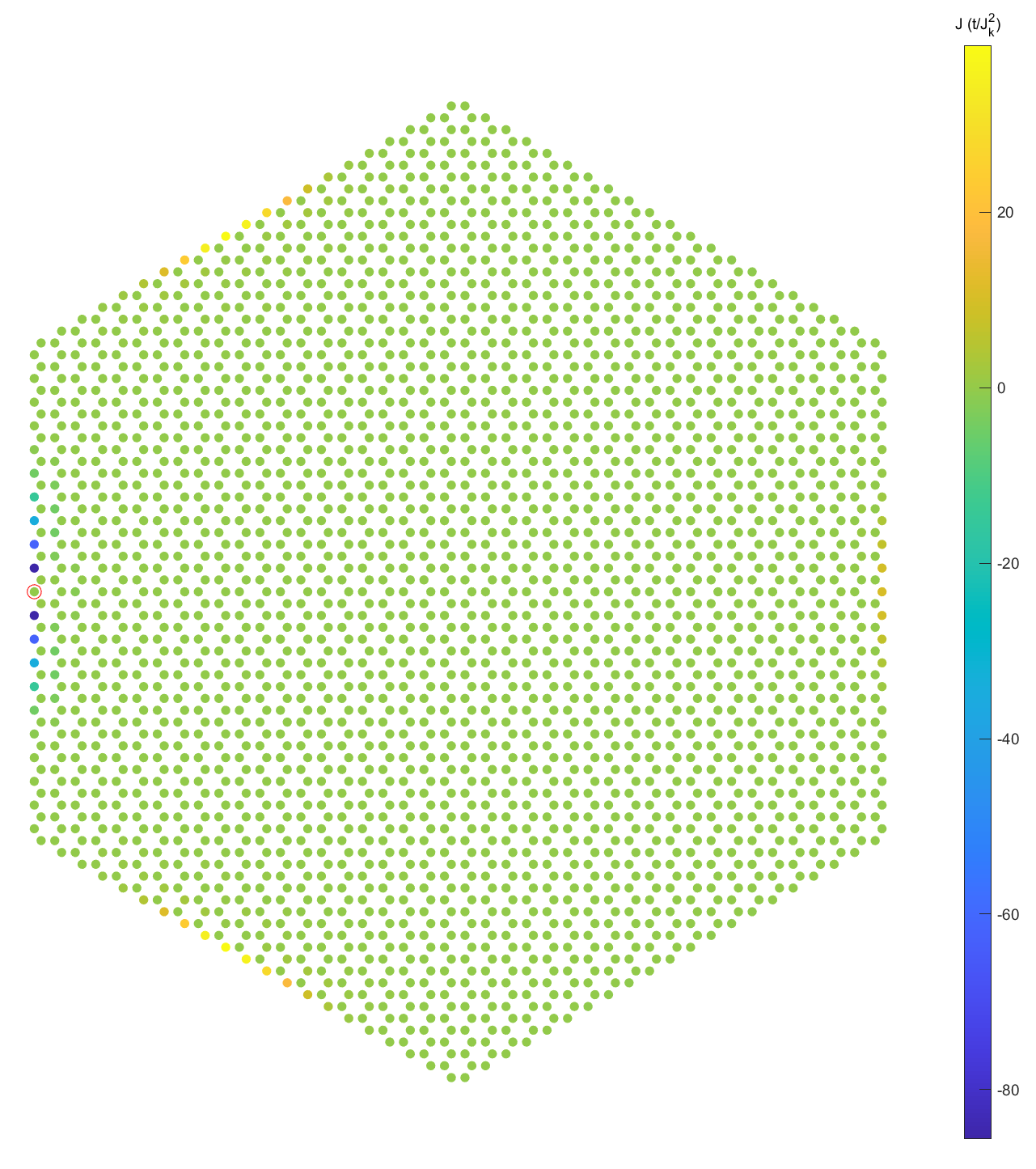}        
    \vspace{-0.8cm}\caption{}
    \end{subfigure}
    
    \centering
       \begin{subfigure}{0.4\textwidth}
   \hspace{-1cm} \includegraphics[width=\linewidth]{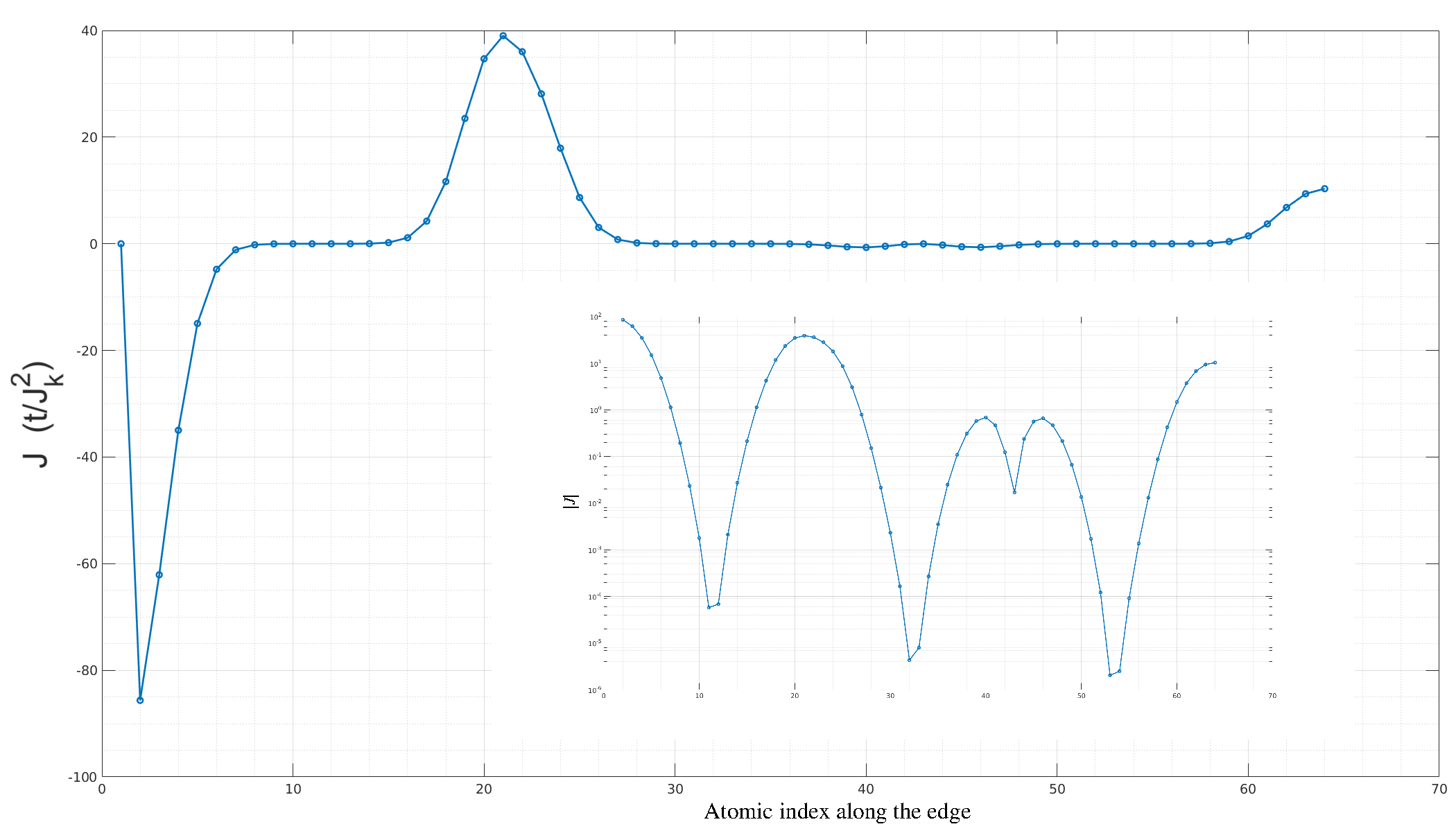} 
    \vspace{-0.2cm}
    \caption{}
    \end{subfigure}  
    
\begin{subfigure}{0.18\textwidth}
   \hspace{-1cm} \includegraphics[width=\linewidth]{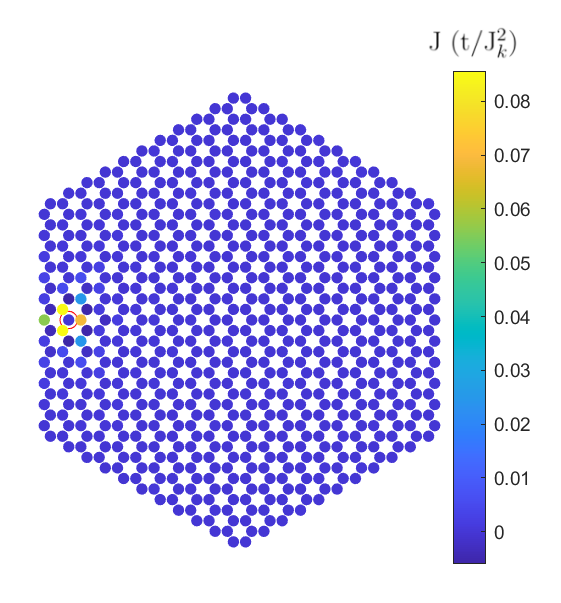}
\vspace{-0.5cm}
    \caption{}
\end{subfigure}
\begin{subfigure}{0.45\linewidth}
    \includegraphics[width=\linewidth]{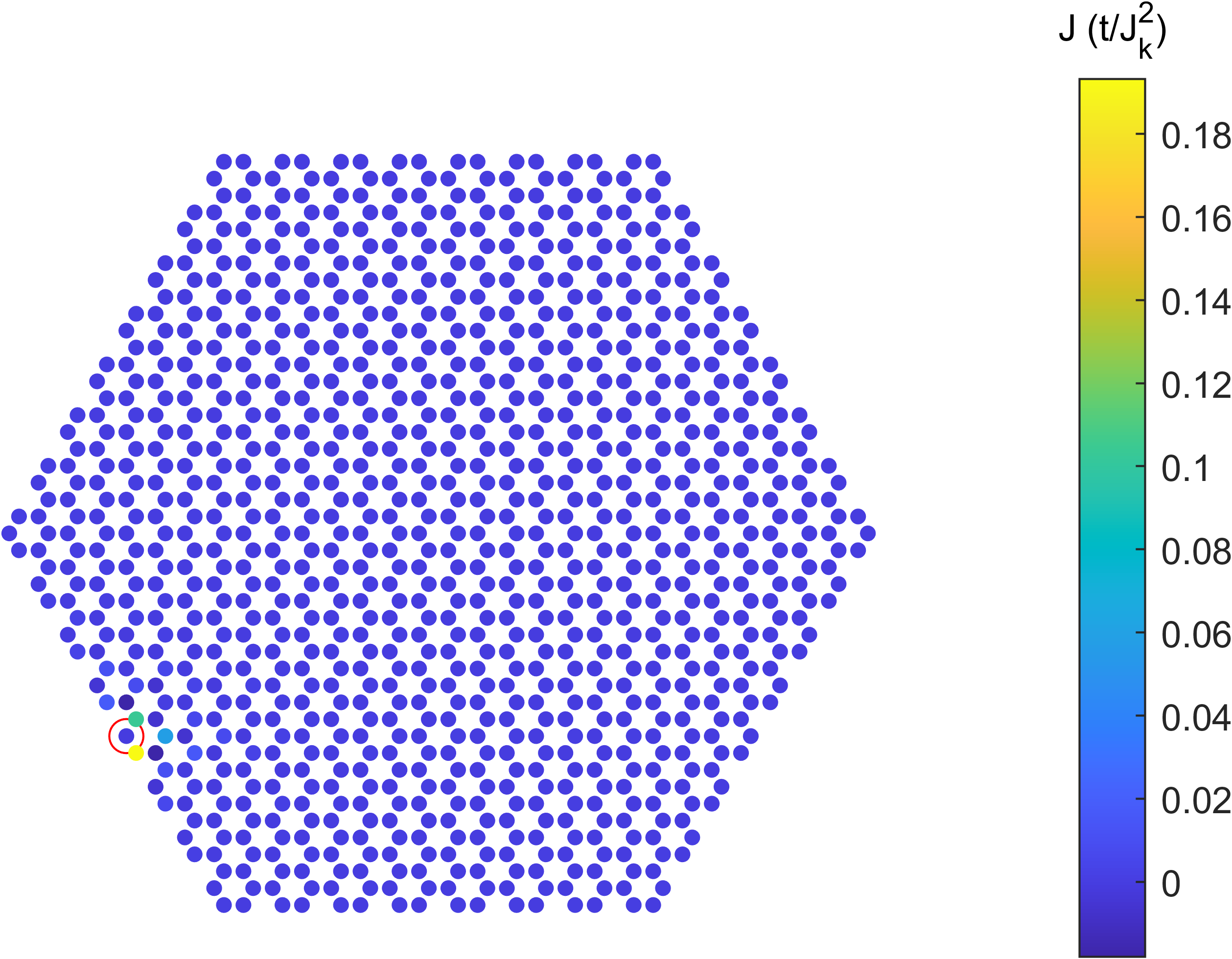}
    \caption{}
\end{subfigure}

\caption{The RKKY interaction in zigzag-edged hexagonal graphene flakes is shown, with respect to the site indicated by a red circle, each side of (a)13 atoms, (b)21 atoms wide. (c)The RKKY interaction is shown along the boundary of zigzag-edged hexagonal flake, when  each side is of 21 atoms wide. In the inset, the absolute value of exchange interaction is shown in log scale. (d)The RKKY interaction is calculated with respect to an atomic site inside the flake, marked by a red circle.(e)The RKKY interaction with respect to an edge atom on an armchair-edged hexagonal flake with 13 edge atoms.}
     \label{fig:hexagonal_zigzag_rkky1}    

\end{figure}

In Fig. \ref{fig:hexagonal_zigzag_rkky1}(a,b) we computed the RKKY interactions at each site within the flake relative to a magnetic moment at the mid-edge site marked by the red circle,  for an edge length of 13, 21 atoms, consisting of 1014, 2646 atoms respectively. In Fig. \ref{fig:hexagonal_zigzag_rkky1}(c) RKKY interaction is shown along the circumference of the flake, with each edge of $N=21$ atoms wide,
featuring an enhancement in exchange coupling between edges of opposite sublattice, where in the inset the absolute value of $J$ is shown in logarithmic scale. 
 The system is neutral, thus the Fermi energy is at the zero energy.  Here each edge belongs to a certain sublattice, alternating between the two sublattice types. Enhancement of the interaction at large distances is exclusive to edges characterized by a different sublattice type in relation to the first impurity marked by the red circle. On the other hand a pronounced ferromagnetic coupling is observed on the same edge as the first moment, both being on the same sublattice.   Conversely, the adjacent edge belongs to the other sublattice and it exhibits antiferromagnetic interactions. Such interactions diminish on the second adjacent edge which is on the same sublattice as the first magnetic moment. On the flake's opposite edge, which is of opposite sublattice type in relation to first magnetic moment, the antiferromagnetic interaction is revived. In pristine bulk graphene, the RKKY interaction decays proportional to the third power of the distance\cite{Sherafati2011}. However here for zigzag-edged hexagonal flakes, it is seen that the interaction persists over extended distances on the flake boundary, attributable to the zero-energy edge states. 

\begin{figure} 
\begin{subfigure}{0.2\textwidth}
    \includegraphics[width=\linewidth]{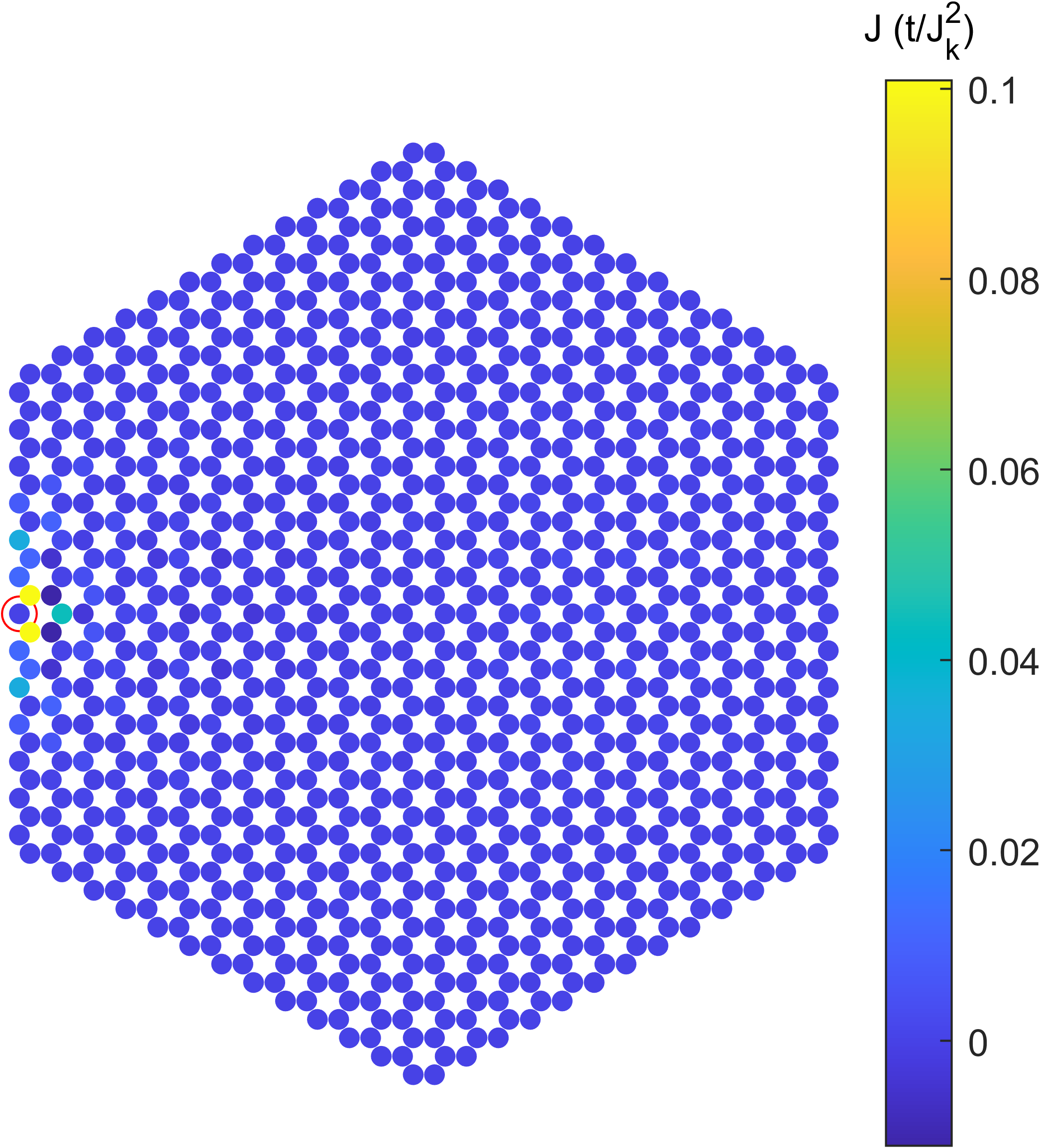}
     \caption{}
\end{subfigure}
\begin{subfigure}{0.25\textwidth}
\centering 
    \includegraphics[width=\linewidth]{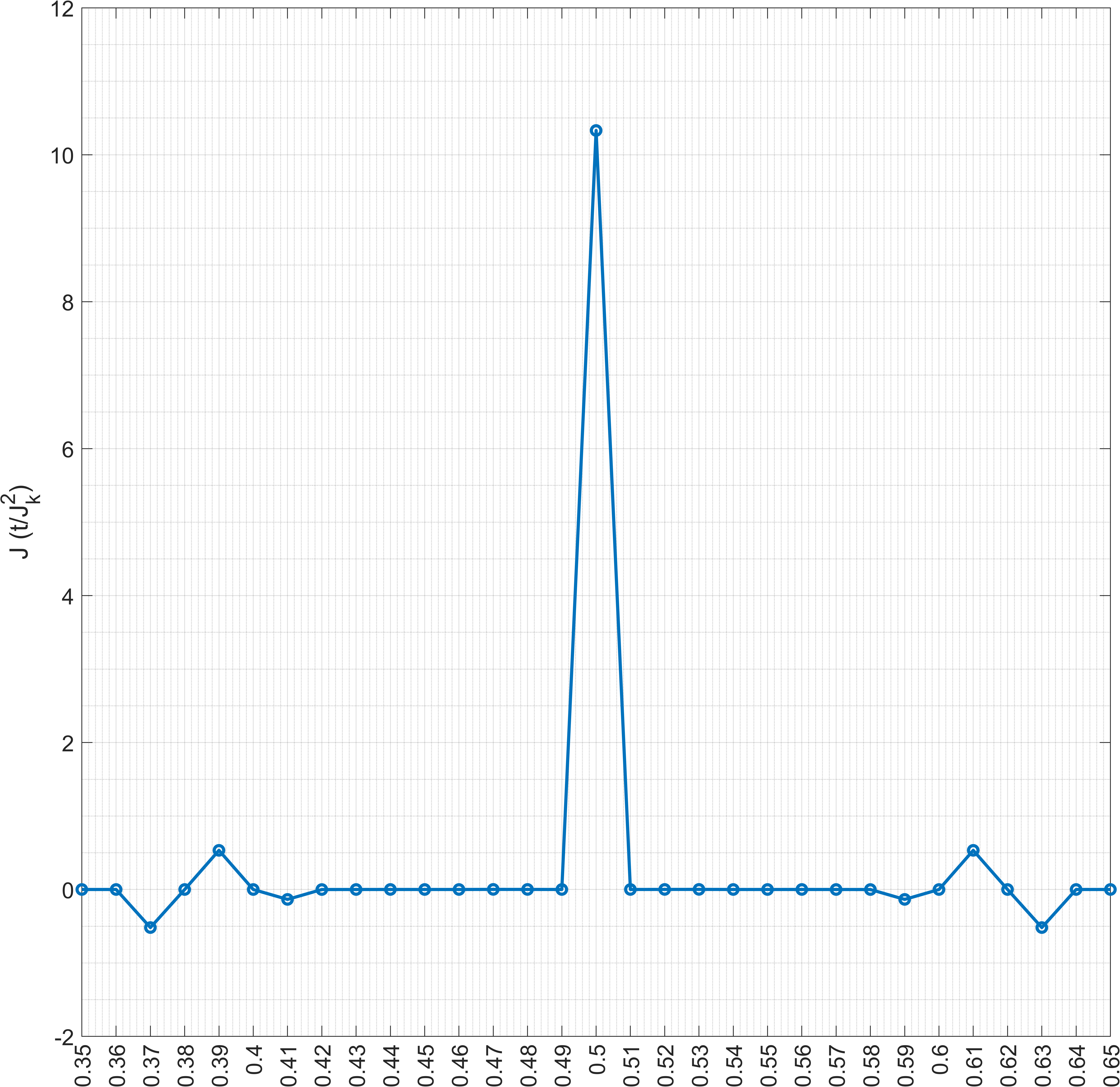}
     \caption{}
     \end{subfigure}
 \caption{(a)The zigzag-edged hexagonal flake with detuned Fermi level is shown. (b)RKKY interaction between the atoms at the midpoint of the opposite edges is shown as a function of electron filling factor, where 0.5 corresponds to neutral graphene.      \label{fig:fermi_level_mismatch} }    
\end{figure}

\begin{figure}[!h]
    \begin{subfigure}{0.4\linewidth}
    \includegraphics[width=\linewidth]{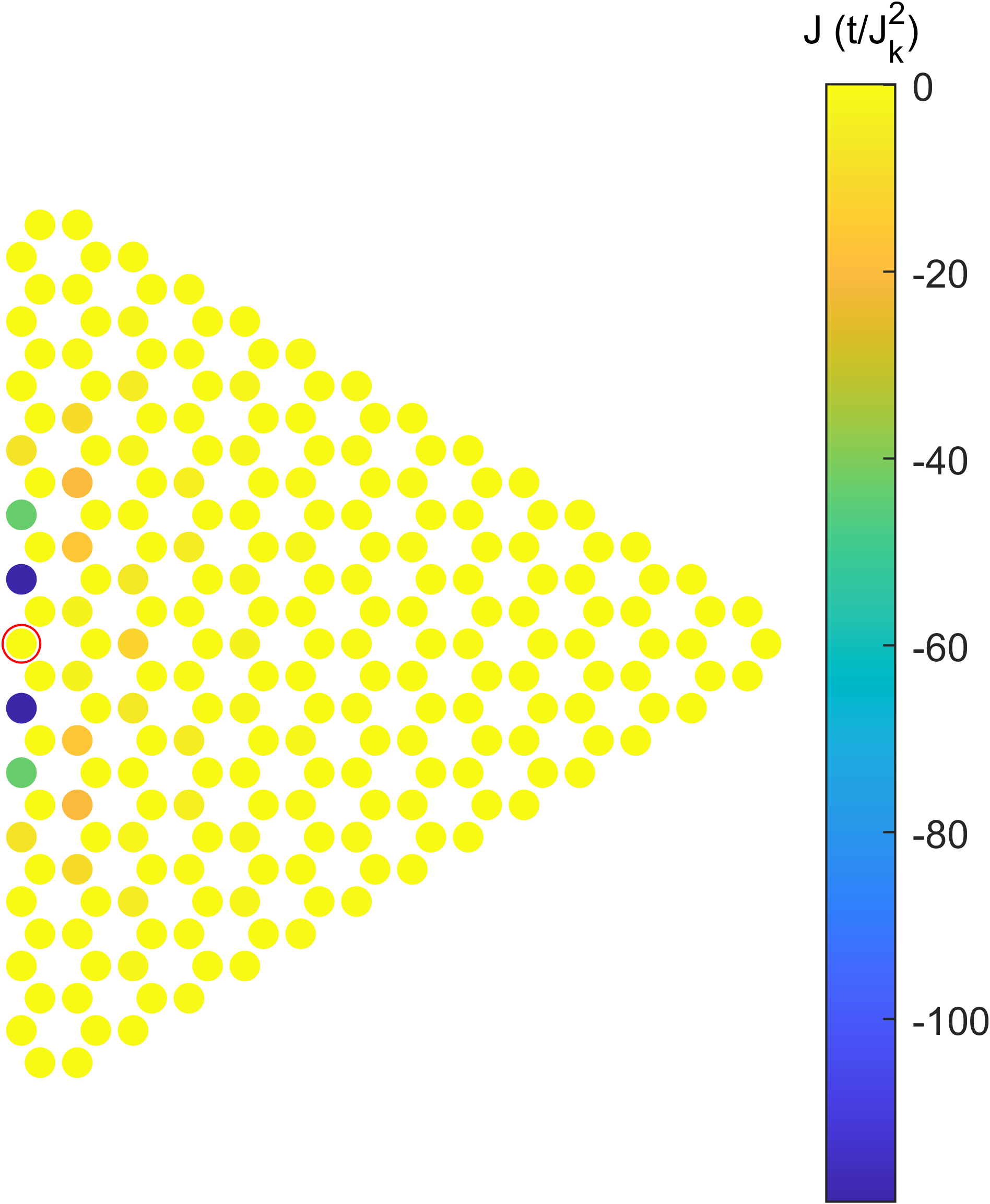}
    \vspace{-0.5cm}
    \caption{}
    \end{subfigure}
        \begin{subfigure}{0.45\linewidth}
        \vspace{0.5cm}
        \includegraphics[width=\linewidth]{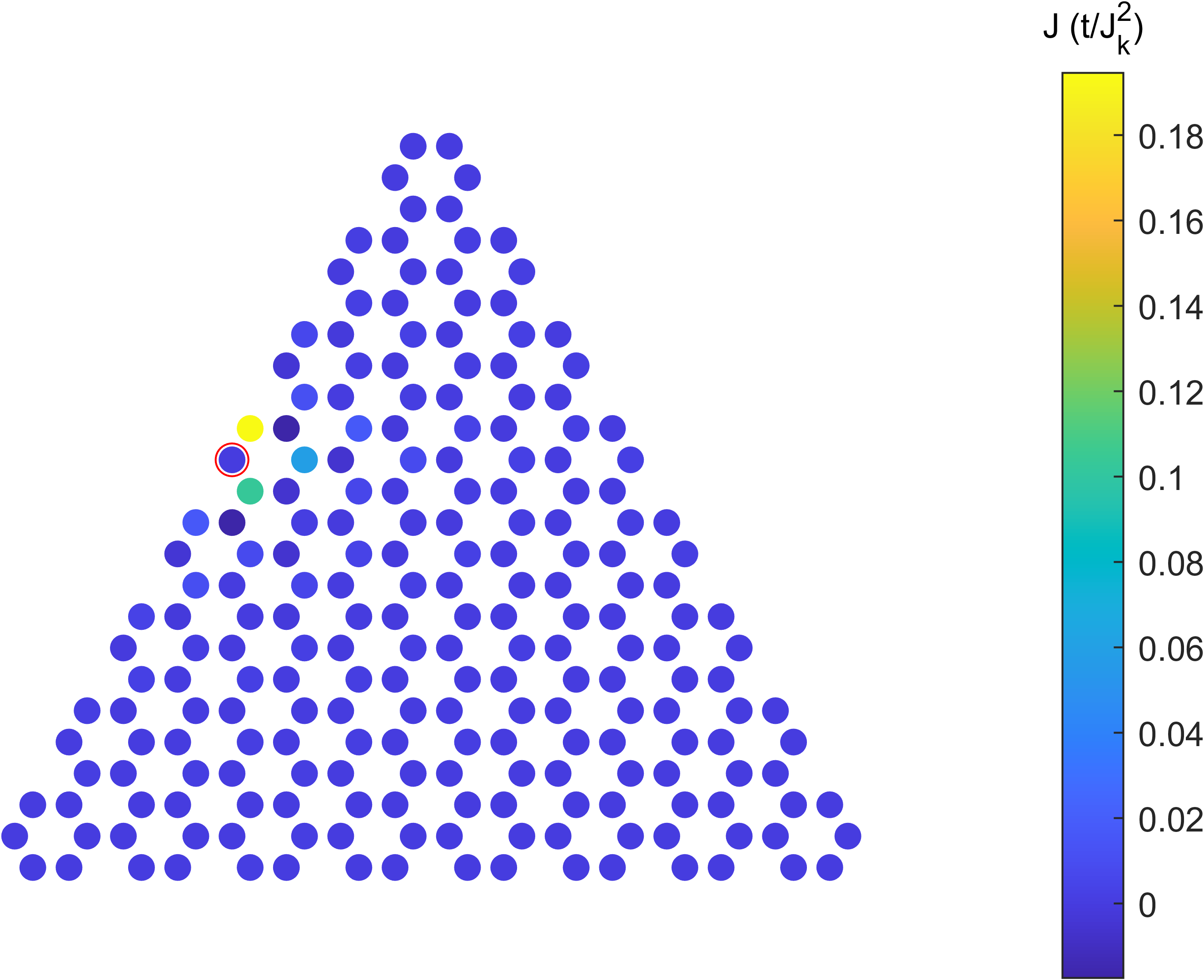}
     \vspace{-0cm}
    \caption{}
    \end{subfigure}     
    \caption{The RKKY interaction across (a)a zigzag-edged triangular graphene flake with 13 edge atoms, with respect to the site highlighted by a red circle, is shown., (b)a triangular graphene flake with $16$ edge atoms   \label{fig:hex_tri}}
\end{figure}

To underscore the significance of zero-energy edge states, we evaluated the interaction for zigzag-edged hexagonal flakes, but with the Fermi level adjusted above the zero energy. Experimentally, doping or applying an external electric field can realize such a modification. The effect of mismatched Fermi level can be seen in Fig. \ref{fig:fermi_level_mismatch}(a) where the system is geometrically the same as the system in Figure \ref{fig:hexagonal_zigzag_rkky1}(a), with the difference in the Fermi level.  Given that the Fermi level does not align with the zero-energy states, the interaction is not enhanced as anticipated.  In Fig. \ref{fig:fermi_level_mismatch}(b) the RKKY interaction between the impurities located at the opposite edges of the zigzag hexagonal flake is shown as a function of filling factor. At half filling, which is the neutral graphene, RKKY interaction becomes maximum.
An alternative way to show the significance of the edge states involves computing the RKKY interaction, not merely at the edge atom, but by positioning it at an interior point of the flake as in figure \ref{fig:hexagonal_zigzag_rkky1}(d). In this scenario, we discern that the RKKY interaction rapidly diminishes, as in bulk, due to the uncoupling from the edge states. In Fig. \ref{fig:hexagonal_zigzag_rkky1}(e) RKKY interaction in an armchair edged hexagonal flake is shown. Here again the first impurity position, shown by the red circle, is fixed whereas the other impurity can be on any site on the flake.  In contrast to zigzag-edged hexagonal flakes an enhancement in RKKY interaction is not seen here neither on the same edge nor the other edges. This can be attributed to the absence of edge states for arm-chair edged flakes, which can be seen in Fig. \ref{fig:hexagonal_eigs}(b) where the zero-energy states are absent. 
 In Fig. \ref{fig:hex_tri}(b) results for an armchair edged triangular flake are presented, here any enhancement in RKKY interaction is absent due to absence of edge states, similar to the armchair edged hexagonal flake.   In Fig. \ref{fig:hex_tri} (c), RKKY interaction is presented for a zigzag-edged triangular flake. Here the presence of zero-energy edge-states does not contribute to interaction enhancement across the adjacent edges. However a strong ferromagnetic coupling is present on the same edge as the first impurity, with the interaction experiencing an abrupt decay across the remaining sites. 

\begin{figure}
    \includegraphics[width=0.8\linewidth]{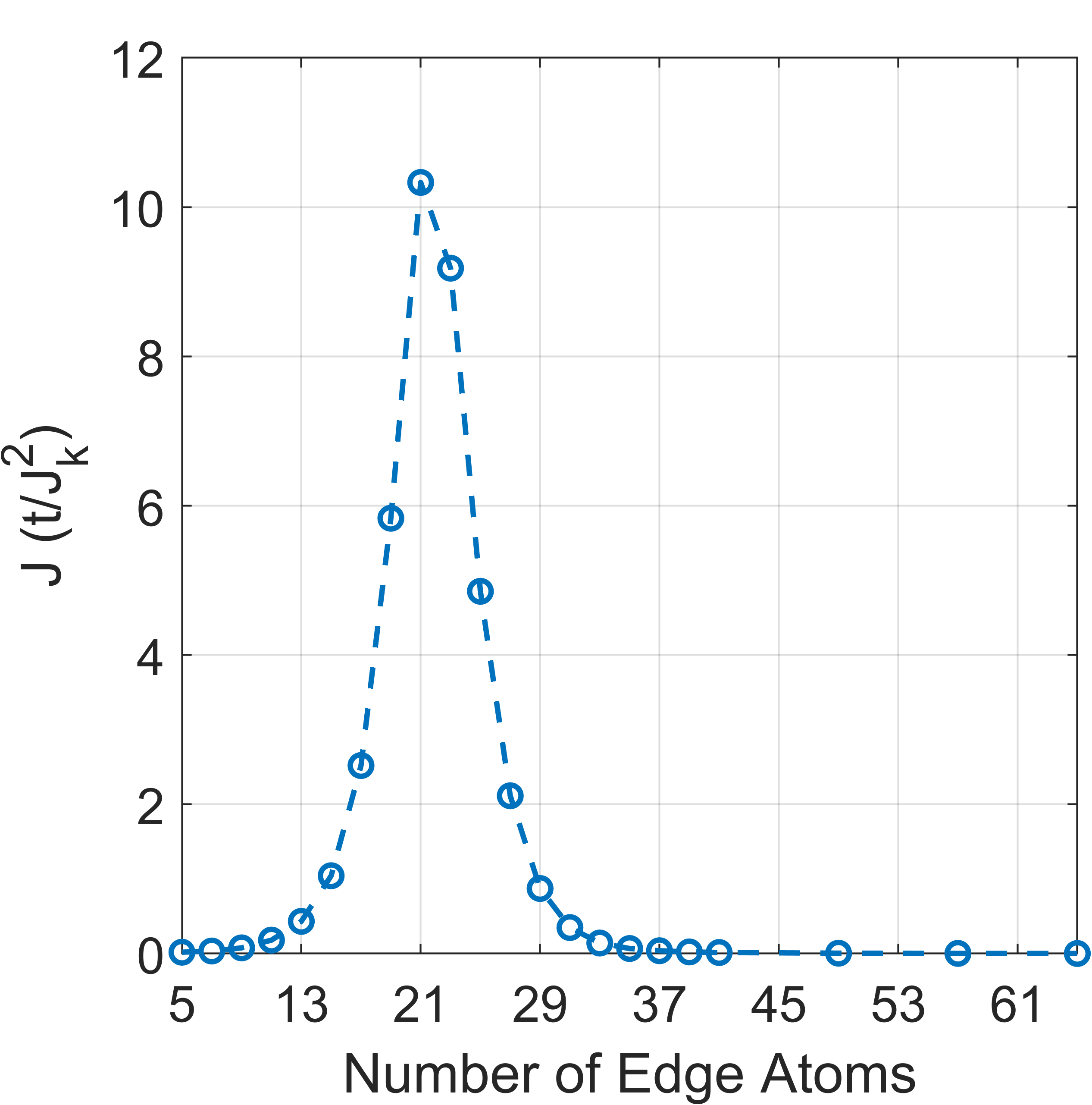}
     \caption{ The figure shows the RKKY interaction strength between midpoints of opposing sites for a zigzag-edged hexagonal flake as a function of system size.}  
     \label{fig:hexagonal_zigzag_rkky}
\end{figure}


\begin{figure}    
    \centering 
    \includegraphics[width=0.8\linewidth]{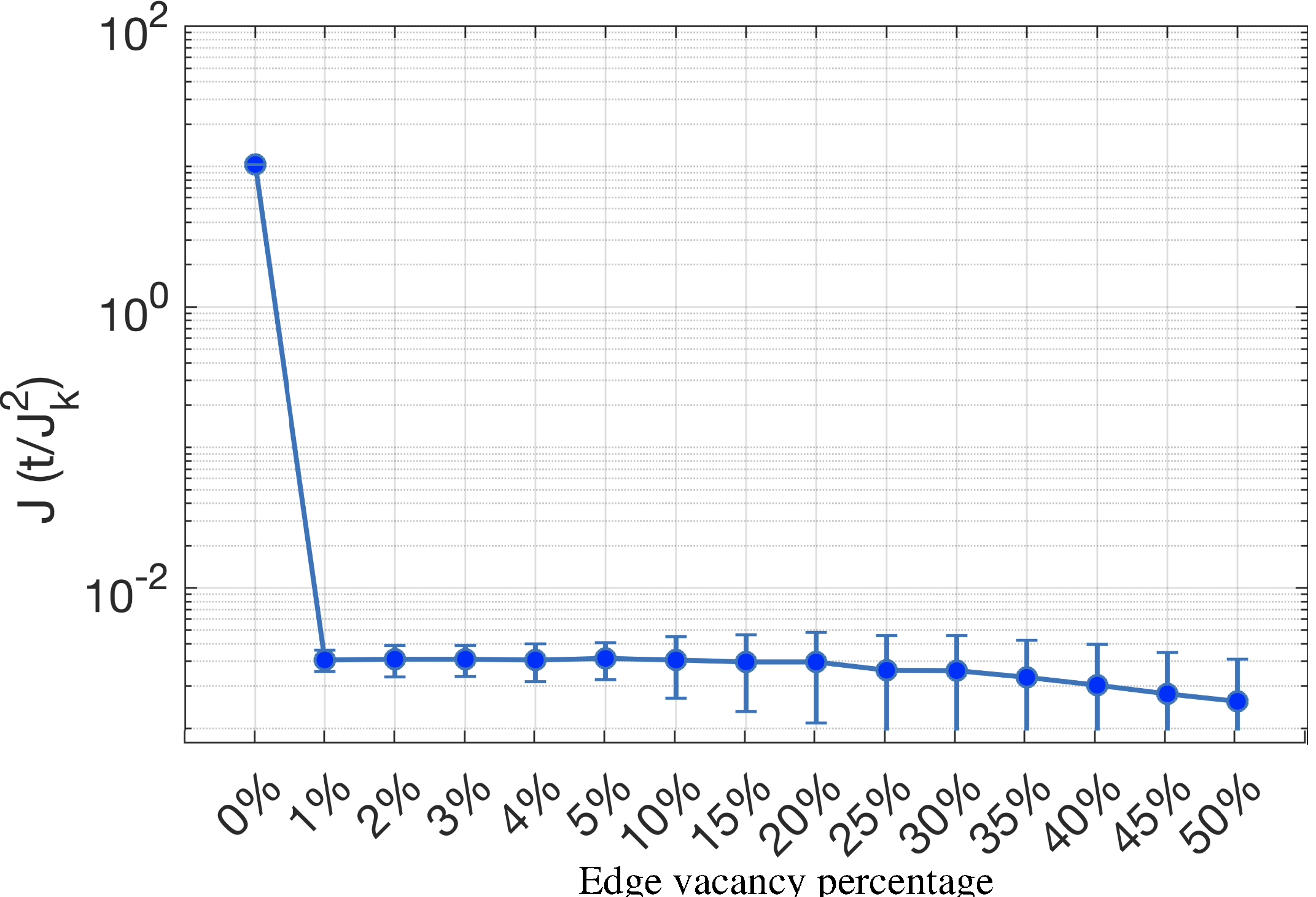}
     \caption{The RKKY interaction is computed for two atoms located at the center of opposing edges, with the presence of varying vacancy percentages. As the number of missing atomic sites increases, as indicated on the x-axis as percentages, it becomes evident that the enhancement of the interaction diminishes.     \label{fig:disorder}}

     \end{figure}

We also elucidate the size dependence of the RKKY interaction within zigzag-edged hexagonal flakes. We computed the interaction strength across varying flake sizes. Our focus remained on the diametrically opposite sides of the flake, gauging the RKKY interaction strength as we vary the size. These findings are represented in Fig. \ref{fig:hexagonal_zigzag_rkky}.  As the size increases, the interaction strength also increases until the flake size reaches edge length of 21 atoms. After this point, the interaction strength starts decreasing again. With increasing flake size, the number of zero-energy edge states also increases, enhancing the coupling.  Nevertheless, when the number of edge atoms exceeds N=21, the increasing inter-atomic distance becomes dominant, leading to a dampening of the interaction. An intricate interplay emerges between the density of states of the zero-energy edge states and the spatial separation of atomic sites. 


Since edge states play an important role in mediating the exchange interaction,  we also studied RKKY interaction for varying percentages of edge vacancies on the flake's boundaries, for zigzag-edged hexagonal flakes. The results are shown in Fig. \ref{fig:disorder}. We considered a hexagonal zigzag-edged system with 13 atoms on each side where two magnetic moments are placed at the midpoints of opposing edges. To simulate the effect of vacancies, we systematically removed atoms from the edges randomly. We sampled the disorder 1000 times for every percentage values and take the average in the end.  As it can be seen in the Fig. \ref{fig:disorder},  the interaction undergoes an abrupt decay even at at 1\% vacancy ratio. This diminishing trend persists as the percentage of vacancies increases. This observation underscores the pivotal role of edge atoms in the the long-range nature 
 of RKKY interaction within zigzag-edged hexagonal flakes.

\section{Results and Discussion}
In this study we have studied RKKY interactions in graphene nanoflakes and unveiled the role of edge states. In particular in zigzag-edged hexagonal graphene nanoflakes edge states lead to long-range interaction between the edges of opposite sublattice type, featuring a pronounced enhancement.  We have also observed an interesting behavior of inter-edge RKKY interactions with the size of nanoflakes, attaining a maximum value when the edges are 21-atoms wide, an effect arising from the interplay of DOS and the decay of RKKY interactions with distance. We also studied the influence of edge vacancies on the RKKY interactions and we have seen that  the magnetic interactions feature a sudden decay with increasing number of edge vacancies.

\begin{acknowledgments}
This work has been supported by the T\"{u}rkiye Bilimsel ve Teknolojik Ara\c{s}t{\i}rma Kurumu (TÜB\.{I}TAK) under Grant No. 115F408.
\end{acknowledgments}

\bibliography{apssamp}

\end{document}